\documentclass[sigconf]{acmart}

\AtBeginDocument{%
  \providecommand\BibTeX{{%
    \normalfont B\kern-0.5em{\scshape i\kern-0.25em b}\kern-0.8em\TeX}}}

\setcopyright{acmcopyright}
\copyrightyear{2022}
\acmYear{2022}
\acmDOI{XXXXXXX.XXXXXXX}

\acmConference[TheWebConf'22]{Make sure to enter the correct
  conference title from your rights confirmation emai}{April 25--29,
  2022}{Online}
\acmPrice{15.00}
\acmISBN{978-1-4503-XXXX-X/18/06}

\begin{document}

\title{Telegram Monitor: Monitoring Brazilian Political Groups and Channels on Telegram}

\author{Manoel Júnior}
\email{manoelrmj@dcc.ufmg.br}
\affiliation{%
  \institution{Universidade Federal de Minas Gerais}
  \city{Belo Horizonte}
  \state{Minas Gerais}
  \country{Brazil}
}

\author{Philipe Melo}
\email{philipe@dcc.ufmg.br}
\affiliation{
  \institution{Universidade Federal de Minas Gerais}
  \city{Belo Horizonte}
  \state{Minas Gerais}
  \country{Brazil}
}

\author{Daniel Kansaon}
\email{daniel.kansaon@dcc.ufmg.br}
\affiliation{
  \institution{Universidade Federal de Minas Gerais}
  \city{Belo Horizonte}
  \state{Minas Gerais}
  \country{Brazil}
}

\author{Vitor Mafra}
\email{vitor.mafra@dcc.ufmg.br}
\affiliation{
  \institution{Universidade Federal de Minas Gerais}
  \city{Belo Horizonte}
  \state{Minas Gerais}
  \country{Brazil}
}

\author{Kaio Sá}
\email{kaiosa@dcc.ufmg.br}
\affiliation{
  \institution{Universidade Federal de Minas Gerais}
  \city{Belo Horizonte}
  \state{Minas Gerais}
  \country{Brazil}
}

\author{Fabrício Benevenuto}
\email{fabricio@dcc.ufmg.br}
\affiliation{
  \institution{Universidade Federal de Minas Gerais}
  \city{Belo Horizonte}
  \state{Minas Gerais}
  \country{Brazil}
}


\begin{abstract}
    Instant messaging platforms such as Telegram became one of the main means of communication used by people all over the world. 
    Most of them are home of several groups and channels that connect thousands of people focused on political topics. However, they have suffered with misinformation campaigns with a direct impact on electoral processes around the world. While some platforms, such as WhatsApp, took restrictive policies and measures to attenuate the issues arising from the abuse of their systems, others have emerged as alternatives, presenting little or no restrictions on content moderation or actions in combating misinformation.
    Telegram is one of those systems, which has been attracting more users and gaining popularity. 
    In this work, we present the ``Telegram Monitor'', a web-based system that monitors the political debate in this environment and enables the analysis of the most shared content in multiple channels and public groups. 
    Our system aims to allow journalists, researchers, and fact-checking agencies to identify trending conspiracy theories, misinformation campaigns, or simply to monitor the political debate in this space along the 2022 Brazilian elections. We hope our system can assist the combat of misinformation spreading through Telegram in Brazil.
    

\end{abstract}



\maketitle

\begin{figure*}[t]
  \centering
  \includegraphics[width=0.84\textwidth]{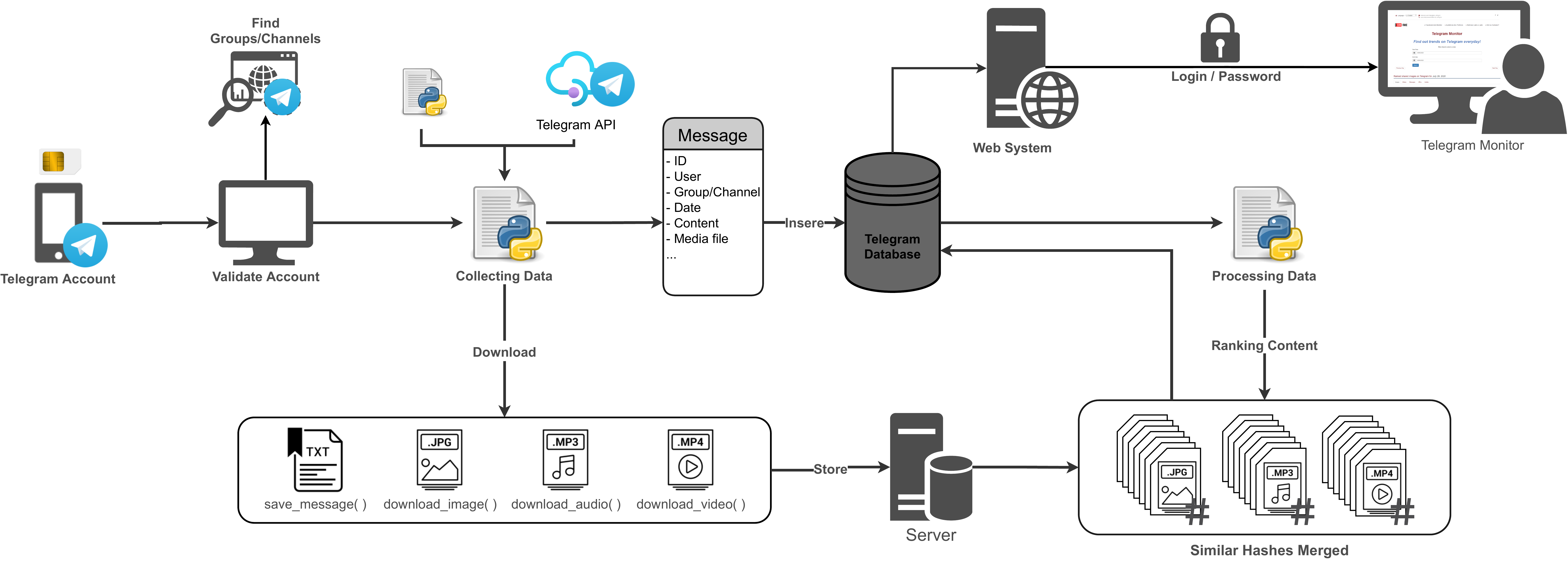}
  \vspace{-0.3cm}
  \caption{General framework architecture of Telegram Monitor.}
  \Description{Framework of Telegram Monitor.}
  \label{fig:architecture}
\end{figure*}

\section{Introduction}

Instant messaging platforms have become one of the main ways to communicate with close friends and also find communities of people who share the same interest, creating large public groups and channels, gathering thousands of users around a topic of discussion.
Telegram ranks among one of the most popular with 400 million\footnote{\url{https://www.statista.com/statistics/234038/telegram-messenger-mau-users/}} monthly active users around the world and has seen a huge growth in its user base. Together with their biggest competitor, WhatsApp, these platforms operate a network that exchanges billions of messages every day\footnote{\url{https://blog.whatsapp.com/connecting-one-billion-users-every-day}}. 
Unfortunately, with the formation of these communities, instant messaging apps also attracted the attention for the spreading of misinformation campaigns and fake news during electoral campaigns~\cite{Resende-WWW2019}. 
These problems, along with public pressure to companies such as Facebook, motivated the establishment of measures to make it difficult to spread content quickly, such as limiting the forwarding of messages~\cite{melo2019can} and banning accounts that violate the platform's terms of use. In this scenario, other messaging platforms, including Telegram itself, end up emerging as an unmoderated alternative in which users are free to discuss whatever they want without any restrictions. 

Although Telegram gained some prominence with a large user base, especially in Brazil, it has been criticized for spreading terrorist content \cite{Telegram-Terrorists}, and hate speech \cite{Telegram-White-Supremacists}, and little has been investigated about this service.
Unlike other social media, it is a hard task to have any information about the huge content circulating on this platform. Due to the closed nature of these systems, not even the most discussed trends or topics are known to those who intend to analyze this environment.

This work is built on the creation of a Telegram public group monitoring system -- the \textbf{Telegram Monitor}. With this system, we seek to help better understand the ecosystem of messaging applications in Brazil, creating an interface that brings out some of the most shared media and messages in political groups and public channels on the platform, facilitating the investigation of the content that circulates on this network.
The Telegram Monitor is an online system that displays, in a period of time selected by the user, the most shared media on these days in the investigated public groups, similar to other initiatives with the same proposal, but for WhatsApp~\cite{whatsappmonitor@icwsm19}. This system is updated daily with images, audios, videos, and messages shared, allowing users to discover the most popular topics that are present on the platform given a period of time.
This can help many researchers and journalists that have access to the system to observe how these platforms are being used and easily identify false stories getting viral on its network, working as a tool to fight against misinformation.
The system's architecture integrates an extensive collection of content from public political groups in Brazil, with the storage and processing of all media shared in these communities. To understand the popularity of each content, the collected media are grouped by similarity, based on a \textit{perceptual hash} algorithm that generates a unique identifier for each content, thus being able to aggregate similar content and quantify how many times each one was shared by the users. Finally, this content is processed on a daily basis, ranking each type of media by popularity and displayed in an online access system through a URL with a link to the system: \url{http://blackbird.dcc.ufmg.br/monitor-telegram/brazil/}. Through an account with a username and password, it is then possible to access all the data and browse the days exploring what was shared in the monitored groups and channels.




\section{System Architecture}


Telegram is a cloud-based, cross-platform instant messaging application, where the accounts are validated by a phone number. 
With this service, users can send messages to their contacts and participate in chat groups of up to 200,000 members and unlimited size channels. They can also invite others to their groups/channels by sharing an invitation URL link.
Our methodology uses data collected from these public groups and channels. The system monitors groups and channels for political discussion in Brazil.
Those are operated either by individuals affiliated with political parties, local community leaders, or ordinary users with an interest in the topic, and can be freely accessed by anyone with the invite link.

In this section, we describe all methodology behind the development of the system of Telegram Monitor.
This main architecture is explained through the flowchart in Figure~\ref{fig:architecture}. In this Figure, we have an overview of all the steps performed to build the online system: 
(i) A phone with a valid Telegram account is set. 
(ii) We find a set of relevant groups and channels on the Web to monitor. 
(iii) A script extracts and structures all messages received. 
(iv) Media files (image, video and audio) are downloaded.
(v) Another step is responsible for parsing these files and grouping the messages by similarity.
(vi) Merged messages are saved containing the total numbers of times they were shared. 
(vii) The online system accesses the database and displays it in the interface according to the user's filters. 
Next, we go through each step separately giving details on how they were performed.

\subsection{Data Collection}

The first step in gathering data on Telegram is to define which are the groups and channels of interest to the collection. For this work, we focus on monitoring Brazilian political public groups and channels. 
A common practice for gathering groups in messaging apps is to use keywords along with the group invitation link pattern to identify potential groups and channels of interest for collection~\cite{garimella2018whatapp,whatsappmonitor@icwsm19,Resende-WWW2019,machado2019study,bursztyn2019thousands}. Telegram group invite links follow the pattern \textbf{\url{https://t.me/joinchat/<GroupID>}} or \textbf{\url{https://telegram.me/<GroupID> }}.
In this way, we created a wide list of terms related to the Brazilian political context. The terms in this list were searched together with the invitation URL pattern on search engines (Google) and social media  (Twitter and Facebook).
In addition to this, Telegram has a group search system within the application itself, where terms were also searched to find more groups.

\begin{figure}[t]
  \centering
  \includegraphics[width=0.85\linewidth]{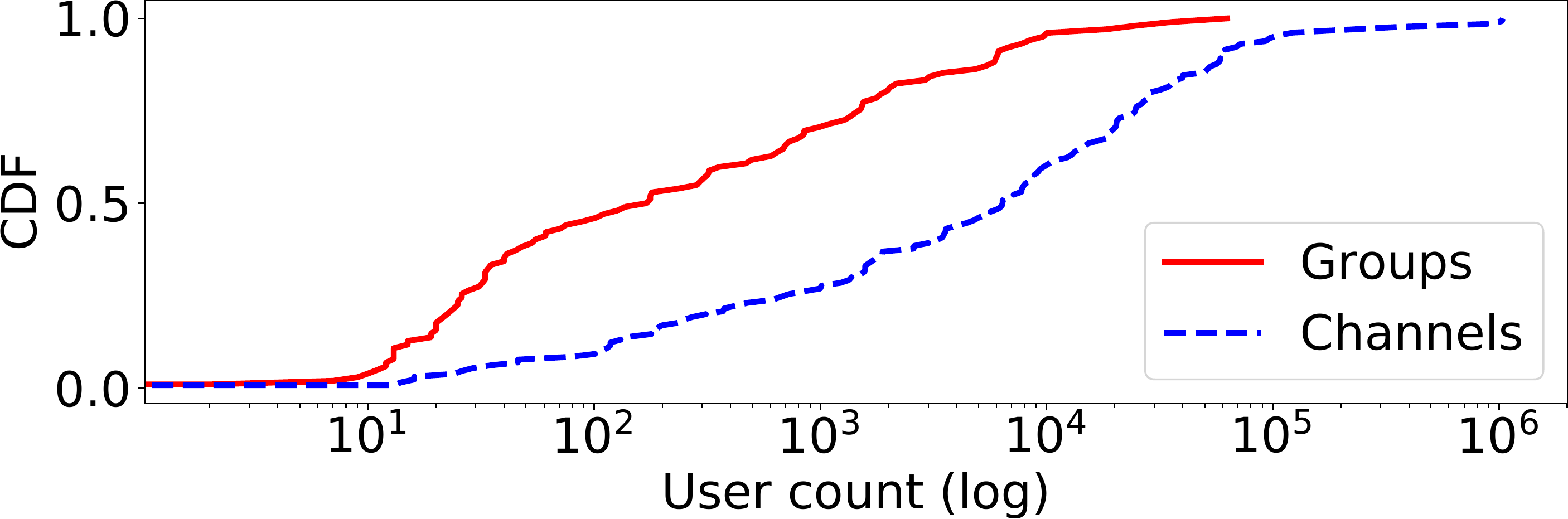}
  \vspace{-0.25cm}
  \caption{Cumulative Distribution Function (CDF) of total of members per groups and channels on Telegram.}
  \label{fig:members}
  \vspace{-0.1cm}
\end{figure}

With this process, we identified and joined a total of 232 public groups and channels related to politics with thousands of members. Some of the selected channels have more than a million members as shown in Figure~\ref{fig:members}. It is interesting to note that around 66\% of the groups and channels found have more than 256 members or subscribers (maximum size of a WhatsApp group). That gives us a hint of how big is the audience that information spreading can reach on those Telegram groups/channels which is much bigger than WhatsApp groups maximum capacity. 

\subsection{Content Processing}

The next step in the development of the system is extracting and processing all the data coming from the groups.
The Telegram platform has an official API\footnote{Telegram API available at: \url{https://core.telegram.org/api}.} in which it is possible to obtain conversation data, in addition to performing actions on behalf of the user, such as joining groups and sending messages programmatically. 
With that, a retroactive collection of data from the conversations was carried out. This data includes text messages and media such as images, videos, audios, documents and any other type of content that the platform supports in its official clients. 


To create a system that allows you to interact with the data and understand the content shared in these groups, it is still necessary to process and organize these messages and media in a way that makes sense to the user who wishes to explore it.
For this, we chose to separate the data by day and, for each day, rank each type of media according to its popularity on the selected date.

\begin{figure}[t]
  \centering
  \includegraphics[width=0.85\linewidth]{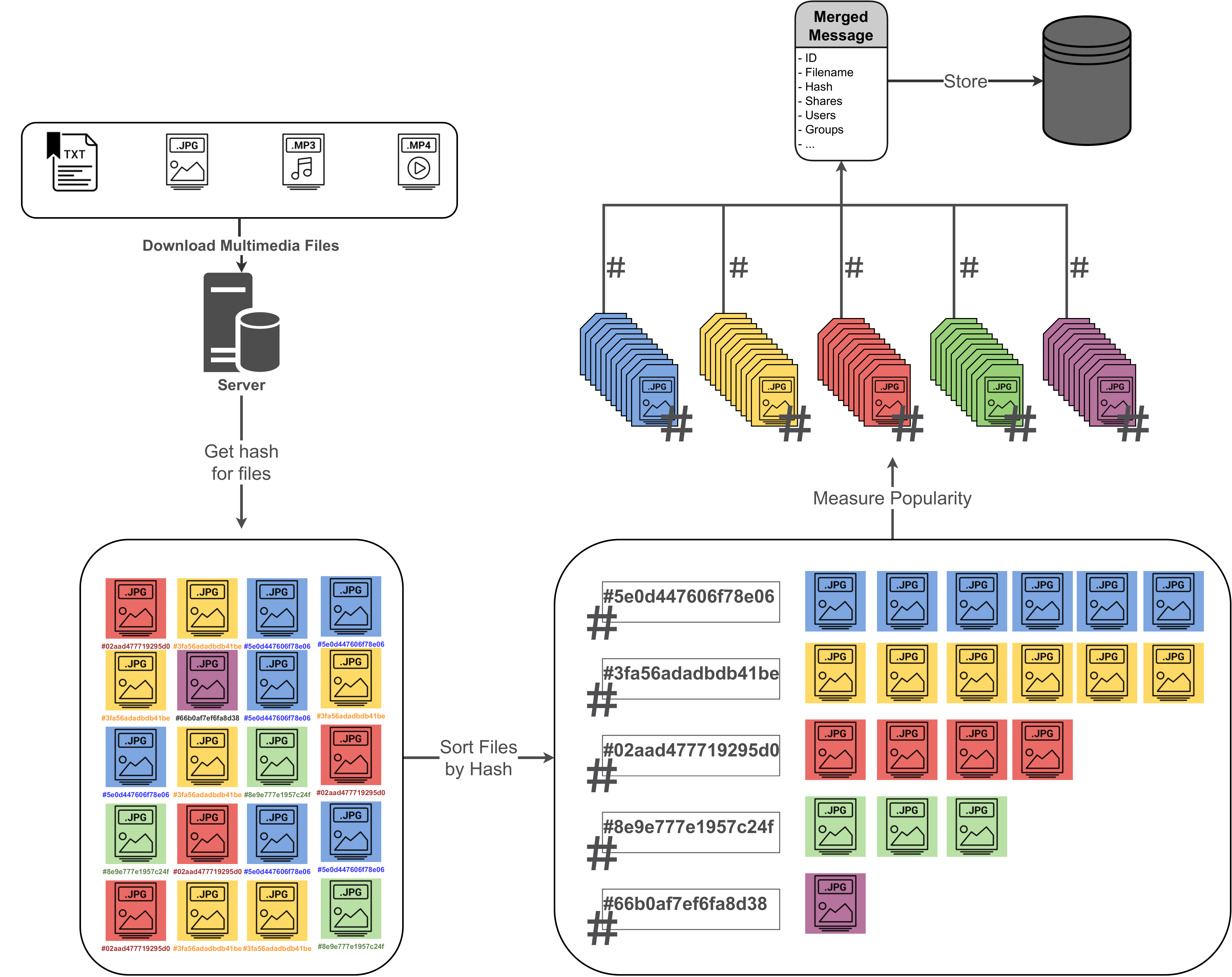}
  \caption{Process to merge files by visual hashes similarity.}
  \label{fig:hashes}
\end{figure}

Here, it is worth noting a crucial step for the functioning of the system: the grouping of similar content. In other social platforms, such as Facebook or Twitter, the data collection of their posts usually already comes with metadata about the number of likes or shares that content gained within the platform. However, with messaging apps like Telegram, each message, image, video, or audio is posted independently. 
Therefore, to define popularity, it is necessary to calculate how many times each media was shared in the period, as this information is not available. 
To be able to track and count how many times the same piece of content has been shared (in other words, measuring their popularity) we need to do the whole process of tracking, grouping, and counting the data, by finding all copies of that message (image, audio or video) among others.
To find copies of the same content, we use the pHash -- \textit{perceptual hashing} algorithm~\citep{monga2006perceptual} on images. This process generates a visual hash that works as a kind of ``fingerprint'' of the content, making it possible to compare two images. For audio and video we also calculate a \textit{hash} from the \textit{checksum} (MD5) that generate a unique value for each different file. Similarly, we also use the Jaccard index to compare text messages and group them.
With those hashes, we can detect content that is identical or very similar (with minor changes) and group them. As a result, it is possible to calculate aggregated information about each piece of content shared, such as determining its popularity, counting in how many groups it appeared and how many different users sent it as shown in Figure~\ref{fig:hashes}.


\subsection{Ethics}

\textit{Telegram Monitor} gathers a considerable amount of data from many Telegram groups and users. To ensure users' privacy, we do not share or disclose any personally identifiable information, such as phone numbers or username.
To prevent misuse, even of aggregated information, we also limit access to our system to a restricted number of people, through an account with a password. 
In addition, they are also informed about the limitations of the data and the potential bias present in our system. As we only use publicly available Telegram groups and official tools provided by the system's API, our data collection does not violate Telegram's terms of service.

\subsection{Online Interface}

\begin{figure}[t]
  \centering
  \includegraphics[width=0.75\linewidth]{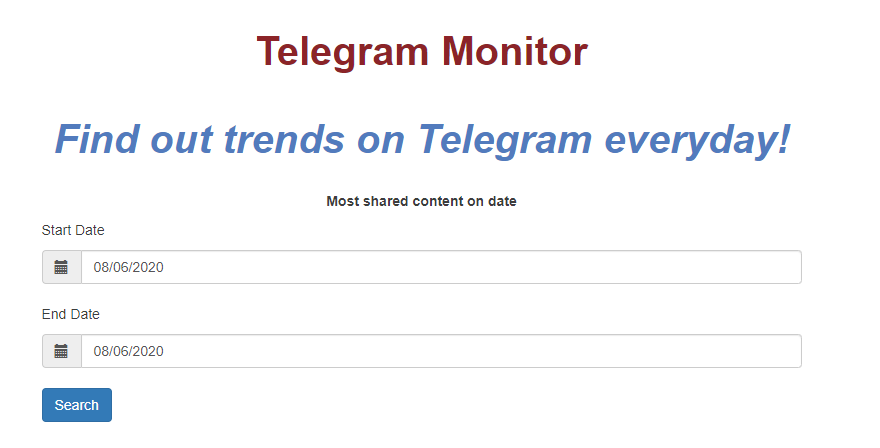}
  \caption{Monitor interface for selecting the period in which one wants to view the content.}
  \label{fig:datapicker}
\end{figure}

\begin{figure}[t]
  \centering
  \includegraphics[width=0.75\linewidth]{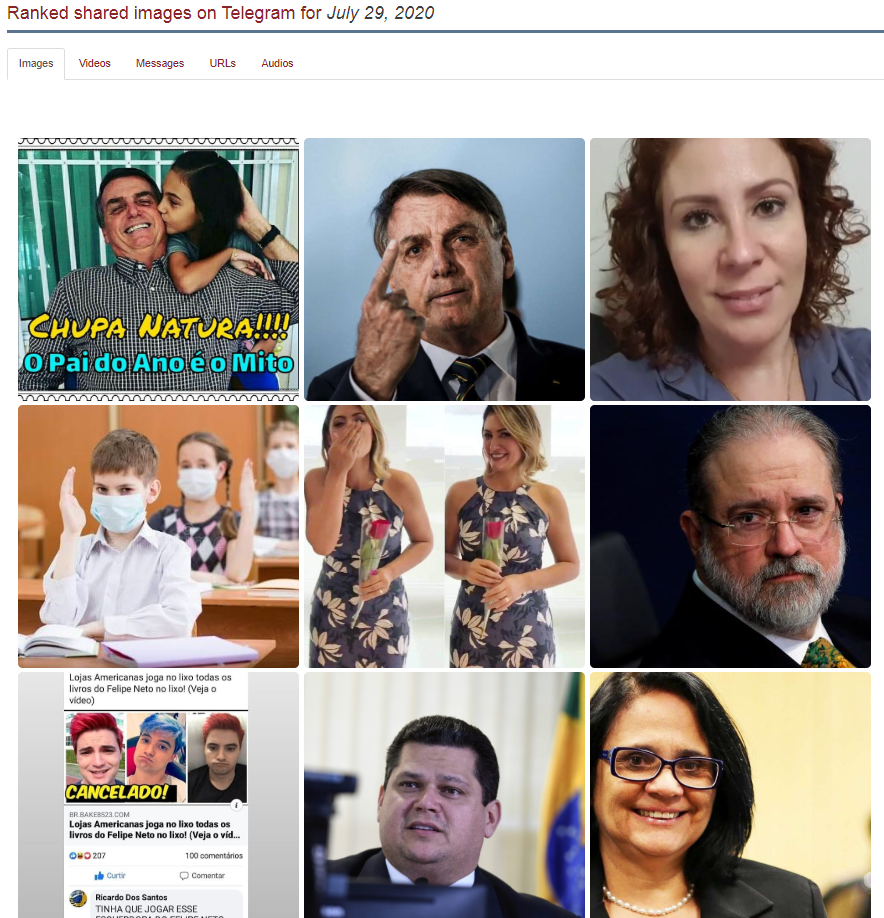}
  \caption{Snapshot of top images of a day in our monitor.}
  \label{fig:homepage}
\end{figure}

After the data is properly separated by date and ranked by popularity, this content is displayed in an online system that fetches the data from the database and displays it to the end user based on the searched period.
The \textit{Telegram Monitor} has restricted access through an account with a username and password on a web page. On this page, the user can select a day (or a period of time with several days), as shown in Figure~\ref{fig:datapicker}, and the system displays a ranking of the most popular content of the selected period.
After picking a date, it is possible to find the most shared content according to the type of media (images, videos, audios, or text). When choosing the media, the content will be ranked based on how many shares it had in that period, showing first those with more shares. An example of the system view can be seen in Figure~\ref{fig:homepage}.

For each content displayed, it is also possible to see more details of it with information about how much it has spread.
In details, we have the total shares, the number of groups in which this content appeared and the number of different users who sent it. It is also displayed the list of names of the groups in which it has been shared. Finally, there is a button that searches for the image on Google to check if that same image appears elsewhere on the Internet.

Since Telegram has such a closed nature, this possibility of navigating through days and seeing what has been popular in groups and channels can help users to better understand that environment, a process that would be hard to execute without such a system.

\subsection{Relevance of the Monitored Groups}

\begin{figure}[t]
  \centering
  \includegraphics[width=0.85\linewidth]{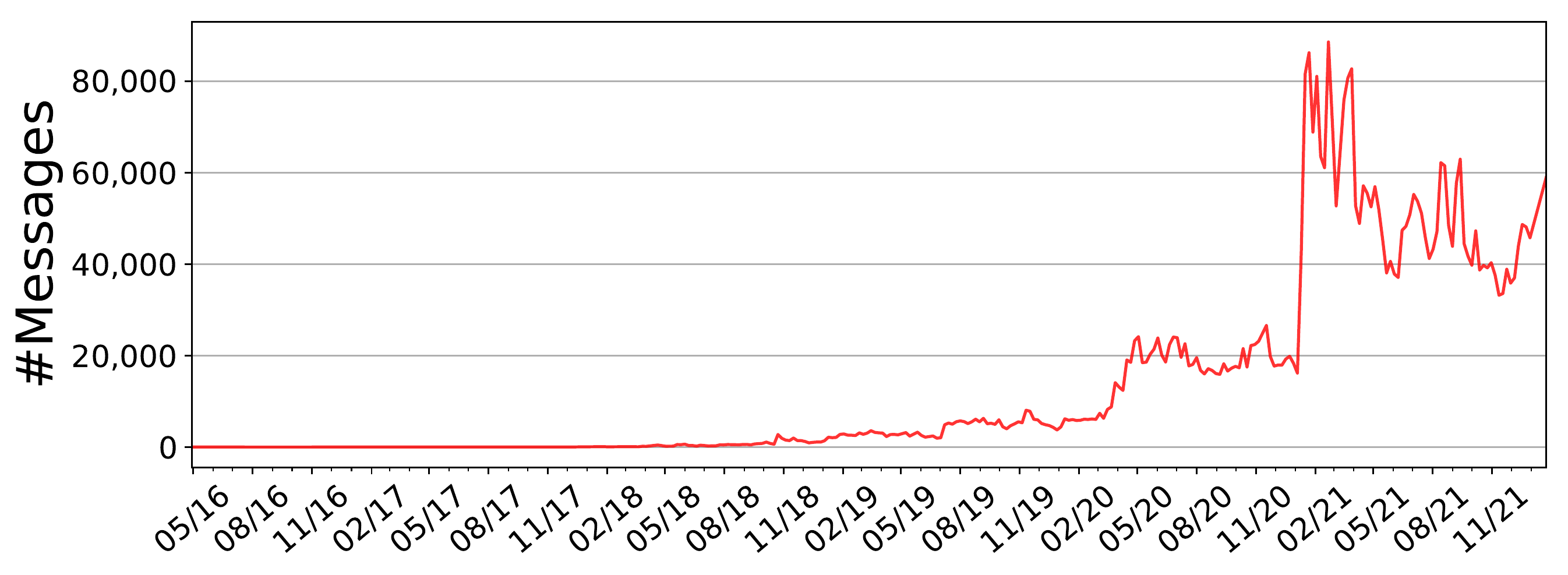}
  \vspace{-0.2cm}
  \caption{Volume of messages sent in monitored groups and channels on Telegram per week.}
  \label{fig:messages}
  \vspace{-0.4cm}
\end{figure}

By joining groups on Telegram, we usually have access not only to messages sent to it after the joining date, but also to the entire history of messages in the groups since their creation. 
Because of that, we can observe the volume of messages exchanged in the groups and channels we monitor since their creation. In Figure~\ref{fig:messages} we show the amount of messages sent per week on those groups and channels. It is possible to notice a relevant increase in the total number of messages in 2020, reaching 20K messages per week and an even bigger increase in early 2021 when that number grew up to four times, jumping to almost 80K messages per week. This result can give us a hint of the importance Telegram has gained, specially in the last year. With groups up to a thousand times larger than WhatsApp, this messaging app has a considerable potential to spread misinformation, although a great part of the content remains distant and away from fact-checking agencies, researchers and journalists, given the difficulty of finding it in such a closed environment.

\section{Social Impact}


Given the upcoming Brazilian elections, scheduled for October 2022, and the serious concerns about the abuse of Telegram by misinformation campaigns, our monitoring system has already attracted some attention as a tool to combat misinformation.

So far, we gave access to our system to a few journalists and fact-checking agencies, who explicitly mentioned it as a data source for their investigation\footnote{\url{https://bit.ly/3umiFMc}}. 
A key challenge for fact-checking organizations and investigative journalism in this space is 
to avoid giving too much attention to conspiracy theories and fake stories that are not getting viral. The concern here is not only to work on something irrelevant, but also to boost the attention to some unpopular misbeliefs.  
Our monitoring system allows fact-checkers to get a sense if a claim is getting viral at least in this specific system. 
In this sense,  our university already became a technical partner of Comprova\footnote{https://projetocomprova.com.br/about/partners/}, a project for collaborative fact-checking in Brazil, recognized by the International Fact-checking network, with more than forty Brazilian media portals. Other key fact-checking agencies in Brazil already showed interest in accessing our system, which we plan to freely provide in the upcoming weeks. 

%

Finally, it is important to mention that Telegram does not have an office or direct representation in Brazil. It has so far ignored Brazilian authorities\footnote{\url{https://bit.ly/34w1Mnk}}, which 
are specially worried about misinformation campaigns questioning the validity of the 2022 elections and the electoral process. Thus, the Superior Electoral Court (TSE), which is responsible for conducting the Brazilian 2022 elections, has also partnered with our project\footnote{\url{https://bit.ly/3oIuRDF}}. 
Our system may represent a way for TSE to get known about possible misinformation campaigns attacking the democrat process through Telegram and take proper actions if necessary. 

\bibliographystyle{ACM-Reference-Format}
\bibliography{references}


\begin{thebibliography}{9}


\ifx \showCODEN    \undefined \def \showCODEN     #1{\unskip}     \fi
\ifx \showDOI      \undefined \def \showDOI       #1{#1}\fi
\ifx \showISBNx    \undefined \def \showISBNx     #1{\unskip}     \fi
\ifx \showISBNxiii \undefined \def \showISBNxiii  #1{\unskip}     \fi
\ifx \showISSN     \undefined \def \showISSN      #1{\unskip}     \fi
\ifx \showLCCN     \undefined \def \showLCCN      #1{\unskip}     \fi
\ifx \shownote     \undefined \def \shownote      #1{#1}          \fi
\ifx \showarticletitle \undefined \def \showarticletitle #1{#1}   \fi
\ifx \showURL      \undefined \def \showURL       {\relax}        \fi
\providecommand\bibfield[2]{#2}
\providecommand\bibinfo[2]{#2}
\providecommand\natexlab[1]{#1}
\providecommand\showeprint[2][]{arXiv:#2}

\bibitem[{Anti-Defamation League}(2019)]%
        {Telegram-White-Supremacists}
\bibfield{author}{\bibinfo{person}{{Anti-Defamation League}}.}
  \bibinfo{year}{2019}\natexlab{}.
\newblock \bibinfo{title}{{Telegram: The Latest Safe Haven for White
  Supremacists}}.
\newblock \bibinfo{howpublished}{ADL - Fighting Hate for Good}.
\newblock
\urldef\tempurl%
\url{https://www.adl.org/blog/telegram-the-latest-safe-haven-for-white-supremacists}
\showURL{%
\tempurl}
\newblock
\shownote{Acessado em 20 de Julho de 2020}.


\bibitem[Bursztyn and Birnbaum(2019)]%
        {bursztyn2019thousands}
\bibfield{author}{\bibinfo{person}{Victor~S Bursztyn} {and}
  \bibinfo{person}{Larry Birnbaum}.} \bibinfo{year}{2019}\natexlab{}.
\newblock \showarticletitle{Thousands of small, constant rallies: A large-scale
  analysis of partisan WhatsApp groups}. In \bibinfo{booktitle}{\emph{2019
  IEEE/ACM International Conference on Advances in Social Networks Analysis and
  Mining (ASONAM)}}. IEEE, \bibinfo{pages}{484--488}.
\newblock


\bibitem[Garimella and Tyson(2018)]%
        {garimella2018whatapp}
\bibfield{author}{\bibinfo{person}{Kiran Garimella} {and}
  \bibinfo{person}{Gareth Tyson}.} \bibinfo{year}{2018}\natexlab{}.
\newblock \showarticletitle{Whatapp doc? a first look at whatsapp public group
  data}. In \bibinfo{booktitle}{\emph{Twelfth International AAAI Conference on
  Web and Social Media}}.
\newblock


\bibitem[Machado et~al\mbox{.}(2019)]%
        {machado2019study}
\bibfield{author}{\bibinfo{person}{Caio Machado}, \bibinfo{person}{Beatriz
  Kira}, \bibinfo{person}{Vidya Narayanan}, \bibinfo{person}{Bence Kollanyi},
  {and} \bibinfo{person}{Philip Howard}.} \bibinfo{year}{2019}\natexlab{}.
\newblock \showarticletitle{A Study of Misinformation in WhatsApp groups with a
  focus on the Brazilian Presidential Elections.}. In
  \bibinfo{booktitle}{\emph{Companion proceedings of the 2019 World Wide Web
  conference}}. \bibinfo{pages}{1013--1019}.
\newblock


\bibitem[Melo et~al\mbox{.}(2019a)]%
        {whatsappmonitor@icwsm19}
\bibfield{author}{\bibinfo{person}{Philipe Melo}, \bibinfo{person}{Johnnatan
  Messias}, \bibinfo{person}{Gustavo Resende}, \bibinfo{person}{Kiran
  Garimella}, \bibinfo{person}{Jussara Almeida}, {and}
  \bibinfo{person}{Fabrício Benevenuto}.} \bibinfo{year}{2019}\natexlab{a}.
\newblock \showarticletitle{{WhatsApp Monitor: A Fact-Checking System for
  WhatsApp}}. In \bibinfo{booktitle}{\emph{Proceedings of the International
  AAAI Conference on Web and Social Media}} \emph{(\bibinfo{series}{ICWSM '19},
  Vol.~\bibinfo{volume}{13})}. \bibinfo{pages}{676--677}.
\newblock


\bibitem[Melo et~al\mbox{.}(2019b)]%
        {melo2019can}
\bibfield{author}{\bibinfo{person}{Philipe Melo},
  \bibinfo{person}{Carolina~Coimbra Vieira}, \bibinfo{person}{Kiran Garimella},
  \bibinfo{person}{Pedro~OS de Melo}, {and} \bibinfo{person}{Fabr{\'\i}cio
  Benevenuto}.} \bibinfo{year}{2019}\natexlab{b}.
\newblock \showarticletitle{Can WhatsApp Counter Misinformation by Limiting
  Message Forwarding?}. In \bibinfo{booktitle}{\emph{Proc. of the Int'l
  Conference on Complex Networks and their Applications (Complex Networks)}}.
\newblock


\bibitem[Monga and Evans(2006)]%
        {monga2006perceptual}
\bibfield{author}{\bibinfo{person}{Vishal Monga} {and}
  \bibinfo{person}{Brian~L. Evans}.} \bibinfo{year}{2006}\natexlab{}.
\newblock \showarticletitle{Perceptual image hashing via feature points:
  performance evaluation and tradeoffs}.
\newblock \bibinfo{journal}{\emph{IEEE Transactions on Image Processing}}
  \bibinfo{volume}{15}, \bibinfo{number}{11} (\bibinfo{year}{2006}),
  \bibinfo{pages}{3452--3465}.
\newblock


\bibitem[Resende et~al\mbox{.}(2019)]%
        {Resende-WWW2019}
\bibfield{author}{\bibinfo{person}{Gustavo Resende}, \bibinfo{person}{Philipe
  Melo}, \bibinfo{person}{Hugo Sousa}, \bibinfo{person}{Johnnatan Messias},
  \bibinfo{person}{Marisa Vasconcelos}, \bibinfo{person}{Jussara Almeida},
  {and} \bibinfo{person}{Fabr\'{\i}cio Benevenuto}.}
  \bibinfo{year}{2019}\natexlab{}.
\newblock \showarticletitle{{(Mis)Information Dissemination in WhatsApp:
  Gathering, Analyzing and Countermeasures}}. In \bibinfo{booktitle}{\emph{The
  World Wide Web Conference}} (San Francisco, CA, USA)
  \emph{(\bibinfo{series}{WWW '19})}. \bibinfo{publisher}{ACM},
  \bibinfo{pages}{818--828}.
\newblock


\bibitem[Tan(2017)]%
        {Telegram-Terrorists}
\bibfield{author}{\bibinfo{person}{Rebecca Tan}.}
  \bibinfo{year}{2017}\natexlab{}.
\newblock \bibinfo{title}{{Terrorists' love for Telegram, explained}}.
\newblock
  \bibinfo{howpublished}{\url{https://www.vox.com/world/2017/6/30/15886506/terrorism-isis-telegram-social-media-russia-pavel-durov-twitter}}.
\newblock


\end{thebibliography}

\end{document}